\documentclass{elsart5-NB}

 \usepackage{graphicx}

\usepackage{amssymb}
\usepackage{amsmath}
\newcommand{\pdagger}{{\phantom{\dagger}}}

\begin{document}

\begin{frontmatter}

\title{Orbital-selective Mott transitions in two-band Hubbard models}

\author[aff1]{Nils Bl\"umer\corauthref{cor1}}
\ead{Nils.Bluemer@uni-mainz.de}
\ead[url]{komet337.physik.uni-mainz.de/Bluemer/}
\corauth[cor1]{}
\author[aff1]{Carsten Knecht}
\author[aff1,aff2]{Krunoslav Po\v{z}gaj\v{c}i\'{c}}
\author[aff1]{Peter G.\ J.\ van Dongen}
\address[aff1]{Institute of Physics, Johannes Gutenberg University, 55128 Mainz, Germany}
\address[aff2]{Institute for Theoretical Physics, J.\ W.\ Goethe University, 60438 Frankfurt/Main, Germany}
\received{7 June 2006}


\begin{abstract}
  The anisotropic two-orbital Hubbard model is investigated at low
  temperatures using high-precision quantum Monte Carlo (QMC)
  simulations within dynamical mean-field theory (DMFT). We
  demonstrate that two distinct orbital-selective Mott transitions
  (OSMTs) occur for a bandwidth ratio of 2 even without spin-flip
  contributions to the Hund exchange, and we quantify numerical errors in
  earlier QMC data which had obscured the second transition.  The
  limit of small inter-orbital coupling is introduced via a new
  generalized Hamiltonian and studied using QMC and Potthoff's
  self-energy functional method, yielding insight into the nature of the
  OSMTs and the non-Fermi-liquid OSM phase and opening the possibility
  for a new quantum-critical point.
\end{abstract}


\end{frontmatter}

  \section{Introduction}\label{Intro}

\vspace{-0.5ex}
  Recently, the suggestion \cite{Anisimov02} that the two consecutive
  phase transitions experimentally observed \cite{Nakatsuji00ab} in
  Ca$_{2-x}$Sr$_{x}$RuO$_4$ should be interpreted as
  ``orbital-selective Mott transitions'' (OSMTs) sparked a cascade of
  related theoretical papers (see \cite{Knecht0506} and references
  therein).

  Microscopic studies of OSMTs usually consider the 2-band Hubbard
  model $H=H_1+H_2+H_3$, where
  \begin{eqnarray*}
  H_1 &=& -\sum_{\langle ij\rangle m\sigma}  t^\pdagger_m c^{\dag}_{im\sigma}
  c^\pdagger_{jm\sigma}\,+\,U\sum_{im} n_{im\uparrow} n_{im\downarrow}\,,\\
  H_2 &=& \sum\nolimits_{i\sigma\sigma'}(U'-\delta^\pdagger_{\sigma \sigma'} J_z)\, n^\pdagger_{i1\sigma}
  n^\pdagger_{i2\sigma'}
  \end{eqnarray*}
  include hopping between nearest-neighbor sites $i,j$ with amplitude
  $t_m$ for orbital $m\!\in\!\{1,2\}$, \emph{intra}- and
  \emph{inter}\-orbital Coulomb repulsion parametrized by $U$ and
  $U'$, respectively, and Ising-type Hund's exchange coupling;
  $n^\pdagger_{im\sigma} =c^{\dag}_{im\sigma} c^\pdagger_{im\sigma}$ for spin
  $\sigma\in\{\uparrow,\downarrow\}$.  In addition,
\[
  H_3=\tfrac{1}{2}J_\perp\sum\nolimits_{im\sigma}c^\dag_{im\sigma}\left(c^\dag_{i\bar{m}\bar{\sigma}}c^\pdagger_{im\bar{\sigma}}
    +c^\dag_{im\bar{\sigma}}
    c^\pdagger_{i\bar{m}\bar{\sigma}}\right)c^\pdagger_{i\bar{m}\sigma}
  \]
  contains spin-flip and pair-hopping terms (with $\bar{1}\equiv 2$,
  $\bar{\uparrow}\equiv \downarrow$ etc.). As in \cite{Knecht0506}, we refer
  to $H_1+H_2+H_3$ with the isotropic coupling $J_z=J_\perp\equiv J$ as
  the $J$-model and to the simplified Hamiltonian $H_1+H_2$ as the
  $J_z$-model; unless noted, $U'=U/2$, $J=U/4$ so that $U'+2J=U$.

  Early theoretical studies suggested that the expected 2 distinct
  OSMTs occur only in the full $J$ model \cite{Koga04a}, but not in
  the Ising type $J_z$-model \cite{Liebsch04,Koga04b}. Thus, it
  seemed as if spin-flip and pair-hopping terms were essential
  ingredients to orbital-selective physics. However, this is not actually
  the case, as
  shown in a low-frequency analysis of high-precision QMC data
  \cite{Knecht0506}  in Fig.~\ref{wImS-spectra}a:
  \begin{figure}[t]
\unitlength0.1\columnwidth
\begin{picture}(10,0)
\put(0,-0.4){\small a)}
\put(5.3,-0.4){\small b)}
\end{picture}
\includegraphics[width=0.48\columnwidth]{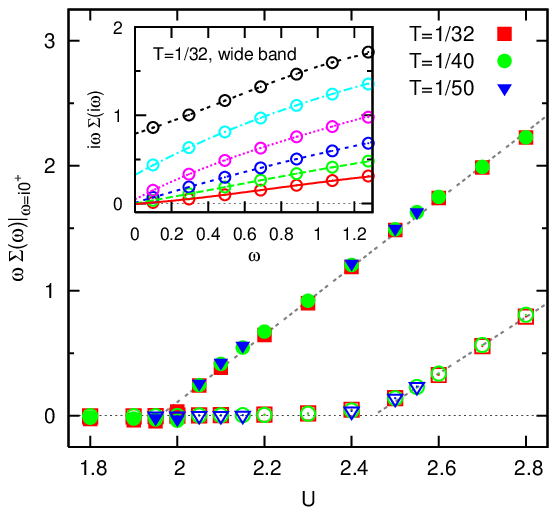}\hfill
\includegraphics[width=0.48\columnwidth]{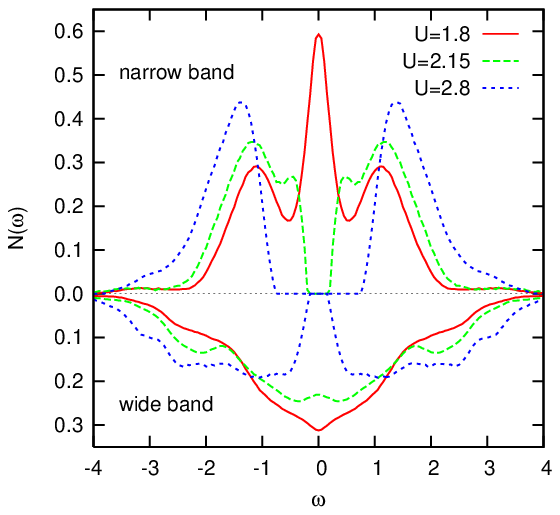}
\caption{a) Low-frequency analysis of self-energy. Main panel: weight
  of singularity at $\omega=0$ for narrow/wide band (filled/open
  symbols) as determined by polynomial fits [shown in inset for
  $U=2.8,2.6,2.4,2.2,2.0,1.8$ (top to bottom)]. b) Spectra for
  metallic, orbital-selective, and insulating phases.}
  	\label{wImS-spectra}
  \end{figure}
  a singularity develops at $\omega=0$ for $U_{c1}\approx 2.0$ in the
  narrow-band self-energy, but only at $U_{c1}\approx 2.5$ in the
  wide-band self-energy (with linear increase for $U>U_{c1/2}$).
  Corresponding spectra (Fig.~\ref{wImS-spectra}b) illustrate the
  characteristics of the 3 distinct phases.

\begin{figure}[t]
\unitlength0.1\columnwidth
\begin{picture}(10,0)
\put(0,-0.4){\small b)}
\put(0,-2.1){\small a)}
\put(5.2,-0.4){\small c)}
\end{picture}
\includegraphics[width=0.48\columnwidth]{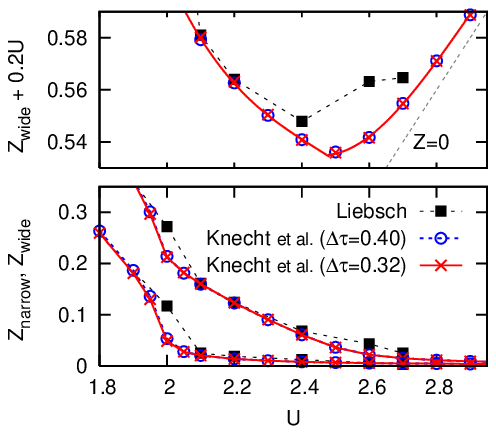}\hfill
\includegraphics[width=0.48\columnwidth]{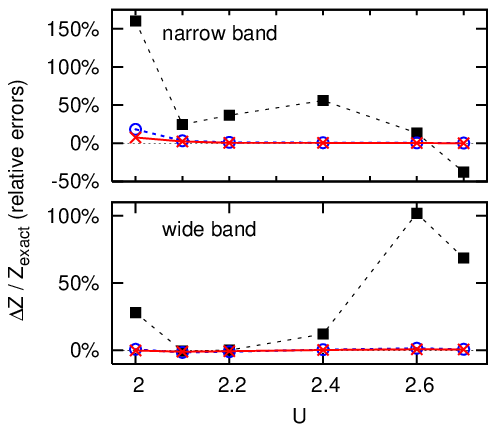}
\caption{a) QMC estimates of quasiparticle weights $Z$ at $T=1/32$: a)
  high-precision data \cite{Knecht0506} (cirles, crosses) clearly shows
  kinks at $U_{c1}\approx 2.0$; b) suitable analysis 
  reveals a second kink at $U_{c2}\approx 2.5$. The second transition
  is lost in the noise of earlier data \cite{Liebsch04}
  (squares) with errors exceeding $100\%$ at both transitions (c).}
  	\label{Z-Liebsch}
  \end{figure}

  Discrete estimates $Z= [1-\text{Im}\,\Sigma(i\pi T)/(\pi T)]^{-1}$
  for the quasiparticle weight clearly show (only) the narrow-band
  transition (Fig.~\ref{Z-Liebsch}a); 
  a second (wide-band) transition
  is visible as kink only in high-precision QMC data (circles,
  crosses) and after adding a linear term (Fig.~\ref{Z-Liebsch}b).
  Possible signals from this transitions cannot be distinguished from
  numerical errors in earlier QMC data \cite{Liebsch04} (squares)
  since they exceed $100\%$ at both transitions
  (Fig.~\ref{Z-Liebsch}c).

  \section{Limit of small inter-orbital coupling}\label{Limit}

\vspace{-0.5ex}
  Clearly, the resolution of
  two orbital-selective Mott transitions with critical interactions
  differing by only about $20\%$ is a very challenging task, in
  particular at temperatures attainable using QMC. In this situation,
  much insight can be gained by abandoning the constraint $U'+2J=U$
  fulfilled in nearly all earlier studies and instead studying the
  limit of small inter-orbital coupling. Hence, we consider
  $H=H_1+\alpha H_2$ with $0\le \alpha \le 1$ so that $\alpha=0$
  corresponds to uncoupled orbitals and $\alpha=1$ to the case studied
  previously.

  It is a priori clear, that for $\alpha=0$ each orbital should
  undergo a usual Mott transition at an interaction determined by the
  corresponding bandwidth ($W=2$ for the narrow, $W=4$ for the wide
  band); note however, that even in this case the QMC results for both
  orbitals have no scaling relation at fixed $T>0$. This is seen in
  Fig.~\ref{alpha}:
  \begin{figure}[t]
\includegraphics[width=\columnwidth]{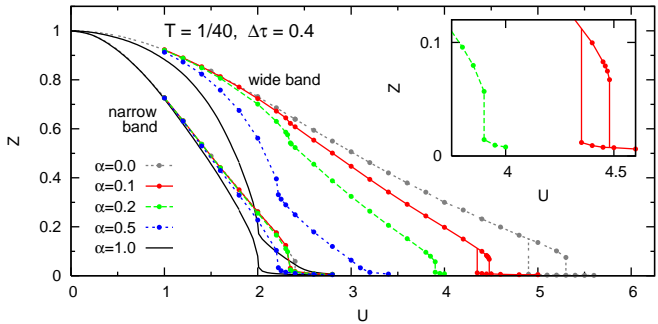}
\caption{Quasiparticle weights for generalized two-orbital model: at
  $T=1/40$, the wide-band transition remains first order (with
  hysteresis) for small enough inter-orbital coupling $\alpha\lesssim
  0.2$.}
  	\label{alpha}
  \end{figure}
  For $\alpha=0$, a large hysteresis region appears in the wide-band
  quasiparticle weight (uppermost, grey line) while only a single
  coexistence point is resolved for the narrow band (at $T=1/40$).
  Very importantly, the wide-band transition evidently remains first
  order at small, but significant inter-orbital coupling
  ($\alpha=0.1$, $\alpha=0.2$). It may be expected that the
  first-order range (in $\alpha$) increases at lower temperatures
  which suggests that the wide-band OSMT might be very weakly first
  order even at $\alpha=1$. However, the alternative of a quantum
  phase transition at some critical value $\alpha=\alpha_c$ is equally
  interesting and warrants further investigation.

\section{Nature of orbital-selective Mott phase}\label{Nature}

\begin{figure}[t]
\unitlength0.1\columnwidth
\begin{picture}(10,0)
\put(0,-0.4){\small a)}
\put(5.3,-0.4){\small b)}
\end{picture}
\includegraphics[width=0.48\columnwidth]{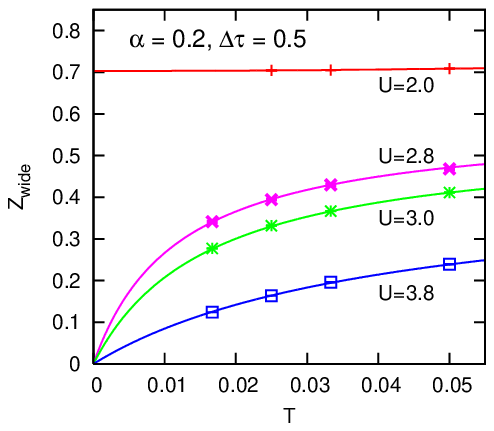}\hfill
\includegraphics[width=0.48\columnwidth]{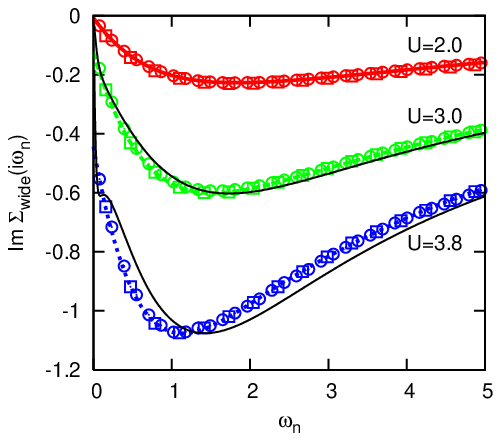}
\caption{a) Discrete QMC estimates of wide-band quasiparticle weights for $\alpha=0.2$.
In OSM phase ($2.3\lesssim U\lesssim 3.9$), $Z_{\text{wide}}$ decays to 0
for $T\to 0$. b) In contrast, the self-energy is practically $T$-independent
in QMC (squares: $T=1/20$, circles: $T=1/40$), consistent with DIA for
$T=0$ (solid lines).}
  	\label{QMC-SFA}
  \end{figure}

\vspace{-0.5ex}
As shown above, the (discretely estimated)
quasiparticle weight $Z$ is not well suited for detecting the second
OSMT. In fact, it is even misleading in the OSM phase: as seen in 
Fig.~\ref{QMC-SFA}a,
  $Z_{\text{wide}}$ is nearly constant as a function of $T$ in the
  metallic Fermi-liquid phase ($U=2.0$). However, this observable
  decays to 0 for $T\to 0$ in the OSM phase. Naively,
  one might conclude that the OSM phase becomes indistinguishable from
  the insulating phase for $T=0$. However, Fig.~\ref{QMC-SFA}b proves
  that this is not the case: the self-energies are practically
  $T$-independent both in the metallic and in the OSM phase [with
  agreement between QMC results for different $T$ and self-energy
  functional theory in dynamical impurity approximation (DIA)]; thus,
  the $T$ dependence in $Z_{\text{wide}}$ is an artifact of the
  discrete approximation.  The non-Fermi-liquid character of the wide
  band in the OSM phase is clearly seen as the finite limit of Im
  $\Sigma$ for $\omega\to i0^+$.

\vspace{1ex}
We acknowledge support by the DFG (FG 559, Bl775/1).
\end{document}